\documentclass[fleqn,twoside]{article}
\usepackage{espcrc2}

\usepackage{graphicx}
\usepackage[figuresright]{rotating}

\newcommand{\AmS}{{\protect\the\textfont2
  A\kern-.1667em\lower.5ex\hbox{M}\kern-.125emS}}

\title{Rare Charm Decays}

\author{S. Pakvasa\address{Department of Physics and Astronomy\\ 
        University of Hawaii, Honolulu, HI  96822}%
     }  
\begin{document}
\begin{abstract}
Rare FCNC decay modes of charmed mesons are reviewed.  The standard model 
predictions, including both short distance and long distance contributions, are 
summarized.  Several new physics options that can give detectable signals are 
described. 
 \vspace{1pc}
\end{abstract}

\maketitle

\section{INTRODUCTION}

The main reason that $D^0 -\bar{D}^0$ mixing and rare (FCNC) decays of charm mesons are interesting is the exciting possibility of new physics beyond the Standard Model(SM) being visible.  In turn this is the case since the GIM mechanism suppressing FCNC transitions is so nearly perfect and the contribution of b quarks is suppressed.  There is potentially a large window for new physics.  The major obstacle in this endeavor is the fact that the long distance effects can be large and furthermore are not always reliably calculable.

What one can do is calculate both short and long distance contributions as reliably as possible in SM and consider possible new physics scenarios with potentially large effects.

\section{STANDARD MODEL}

\subsection{Radiative Decay Modes}

First consider radiative modes of the kind 
$D \rightarrow V \gamma_, \ D_s \rightarrow V \gamma$.  
The short distance effective Hamiltonian for $c \rightarrow u + \gamma$ is given by
\begin{equation}
\quad \quad {\cal{H}}_{\rm{eff}} = - \frac{4G_F}{\sqrt{2}}
\sum_{i=1}^{10} C_i (\mu) O_i (\mu)
\end{equation}
where $c_i$ are the Wilson Coefficients that carry the content 
of the short distance physics inside the loop and $O_i$ are the relevant
operators.  
Matrix elements will contain long distance physics.  
New physics can affect the matching conditions at $\mu=M$. 
In the SM the predicted branching ratio $({\cal B})$ for $c \rightarrow u \gamma$ 
is found to be 
$\sim 10^{-8}$ after several surprising enhancements.  
However, even this is very small compared to the long distance
contributions 
from $s$ and $u$ channel poles from nearby states and VMD contributions
from 
$\rho, w, \phi ..$ These estimates suffer from uncertainties from
several sources.  
The mass $m_c$ is too large for the chiral symmetry approximation and it is
too small for the HQET approximation.  Hence, one has to resort to inspired 
guesswork \cite{burdman,greab,bajc}. In any case, the end results seem
reasonable and vindicated by recent measurements.  The ${\cal B}$ are in the
range of $10^{-5}$ to $10^{-7}$.  The recent measurement of ${\cal B} (D^0
\rightarrow \phi^0 \gamma) \sim 2.6\times 10^{-5}$ by BELLE \cite{tajima} is well within the expected range.

The bottom line is that it is extremely unlikely that new physics in the 
short distance $c \rightarrow u \gamma$ can be separated 
from the much larger SM long distance(LD) contributions.

\subsection{Really Rare Decay Modes}

We now turn to rare decay modes which have a better chance for new
physics to show up \cite{burdman1,fajfer}.  
The decay mode $D^0 \rightarrow \gamma \gamma$ is also long distance
dominated, the largest contribution coming from the on-shell $KK$, $KK^*$  
and $K^*K^*$ intermediate states, with an estimate of $(1-3)\times 10^{-8}$ 
for the ${\cal B}$, to be compared to $3\times 10^{-11}$ 
from a short distance estimate.  The current bound is $3\times 10^{-5}$.

The mode $D^0 \rightarrow \mu^+ \mu^-$ 
has a short distance contribution
to ${\cal B}$ of $10^{-18}$; the LD contribution is 
dominated by on-shell $\gamma \gamma$ 
states and with the above estimate for 
$\gamma \gamma$, one finds a BR
in the 
range $(3-8)\times 10^{-13}$.  The $e^+e^-$ mode in the SM is suppressed by
$(m_e/m_\mu)^4$.

The dilepton modes of the kind $D \rightarrow X \ell^+ \ell^-$ are very
interesting.  
The SD component can be calculated and LD component is 
estimated with the dominant contribution coming from specific 
hadronic states.  Intermediate vector meson states give 
rise to sharp peaks.  The  estimates for the branching ratios are $\sim 10^{-6}$ for 
$D^+ \rightarrow \pi^+ \ell^+ \ell^-$ and $10^{-5}$ for $D_s \rightarrow \pi \ell^+ \ell^-$.

Other dilepton modes which are relatively rare are $D^+ \rightarrow e^+
\nu_e, D_s \rightarrow e^+ \nu_e,$  
and $D^0 \rightarrow l^+ l^- \gamma$ with expected ${\cal B}$ 
of $3\times 10^{-9}, 1.6\times 10^{-7}$ and $10^{-9}$ repectively.

\section{NEW PHYSICS ANALYSIS}

One can consider a wide variety of new physics scenarios which can
contribute to rare 
D decays \cite{burdman1,fajfer,fajfer1,akeroyd}.  For example (i) Supersymmetry, either MSSM or R-parity violating, (ii) new degrees of freedom such as (a) extra scalars, (b) extra gauge bosons, (c) extra fermions, (d) extra dimensions, (iii) new strong dynamics such as varieties of Technicolor, top condensates etc., (iv) others.

Some of these give rise to potentially observable rates for some of the rare charm decay modes.  We review these below.

\subsection{MSSM}

In MSSM there are new Flavor Changing couplings and there are possible large contributions to some rare decay modes.

To calculate R-parity conserving SUSY contributions, one employs the
so-called 
{\it mass insertion approximation}.  

In this approach, a squark propagator becomes 
modified by a mass insertion that changes the squark flavor. 
For convenience, one expands the squark propagator in powers of the 
dimensionless quantity $(\delta^u_{ij})_{\lambda\lambda'}$, 
\begin{equation}
(\delta^u_{ij})_{\lambda\lambda'}={(M^u_{ij})^2_{\lambda\lambda'}
\over M^2_{\tilde q}}\ \ , 
\label{delta}
\end{equation}
where $i\neq j$ are generation indices, $\lambda,\lambda'$ denote the
chirality, $(M^u_{ij})^2$ are the off-diagonal elements of the 
up-type squark mass matrix and $M_{\tilde q}$ represents the average 
squark mass.  The exchange of squarks in loops thus leads to FCNC
through loop diagrams.  

If one works in a basis which diagonalizes the fermion mass 
matrices, then sfermion mass matrices (and thus sfermion 
propagators) will generally be nondiagonal.  As a result, 
flavor changing processes can occur.  One can use 
phenomenology to restrict these FCNC phenomena.  The 
$Q=-1/3$ sector has yielded fairly strong constraints but 
thus far only $D^0$-${\bar D}^0$ mixing has been used to limit 
the $Q=+2/3$ sector.  In our analysis, we have taken charm FCNCs 
to be as large as allowed by the $D$-mixing upper bounds.  


For the decays $D \to X_u \ell^+ \ell^-$, the gluino contributions will 
occur additively relative to those from the SM and so we can 
write for the Wilson coefficients, 
\begin{equation}
C_i = C_i^{\rm (SM)} + C_i^{\rm \tilde g}.
\label{wilson}
\end{equation}
To get some feeling for dependence on the 
$(\delta^u_{12})_{\lambda\lambda'}$ parameters, we display 
the examples 
\begin{equation}
C_7^{\rm \tilde g} \ \propto \ (\delta^u_{12})_{\rm LL}\ {\rm and} \ 
(\delta^u_{12})_{\rm LR}, 
\qquad 
C_9^{\rm \tilde g} \ \propto \ (\delta^u_{12})_{\rm LL}, 
\label{right}
\end{equation}
whereas for quark helicities opposite  
to those in the operators of Eq.~(\ref{right}), one finds 
\begin{equation}
{\hat C}_7^{\rm \tilde g} \ \propto \ (\delta^u_{12})_{\rm RR}\ 
{\rm and} \ (\delta^u_{12})_{\rm LR}, 
\qquad 
{\hat C}_9^{\rm \tilde g} \ \propto \ (\delta^u_{12})_{\rm RR}.
\label{wrong} 
\end{equation}
Moreover, the 
term in ${\hat C}_7^{\rm \tilde g}$ which contains 
$(\delta^u_{12})_{\rm LR}$ experiences the enhancement factor 
$M_{\tilde g}/m_c$. 

For some modes in $D \to X_u \ell^+ \ell^-$, the effect of the 
squark-gluino contributions can be large relative to the SM 
component, both in the total branching ratio and for certain 
kinematic regions of the dilepton mass.  The mode 
$D^0 \to \rho^0 e^+ e^-$ is especially interesting since the ${\cal B}$ can be as high as $1.3\times 10^{-5}$ (only a factor of 10 below the current bound) and the enhancement is in the low $m^2_{ee}$ region.

\subsection{R-parity violating SUSY}
The effect of assuming that $R$-parity can be violated is to 
allow additional interactions between particles 
and sparticles.  Ignoring bilinear terms which are 
not relevant to our discussion of FCNC effects, we introduce 
the $R$-parity violating (RPV) super-potential of trilinear couplings, 
\begin{eqnarray}
{\cal W}_{{\not R}_p} & = & \epsilon_{ab} [ 
1/2 \lambda_{ijk}L^a_iL^b_j\bar{E}_k
+\lambda'_{ijk}L_i^aQ^b_j\bar{D}_k  \\ \nonumber
 & + & 
  1/2  \epsilon_{\alpha\beta\gamma}\lambda^{''}_{ijk}\bar{U}^\alpha_i
\bar{D}^\beta_j\bar{D}^\gamma_k ]
\label{rpv1}
\end{eqnarray}
where $L$, $Q$, $\bar E$, $\bar U$ and $\bar D$ are the standard 
chiral super-fields of the MSSM and $i,j,k$ are generation indices.
The quantities $\lambda_{ijk}$, $\lambda_{ijk}'$ and 
$\lambda^{''}_{ijk}$ are {\it a priori} arbitrary couplings 
which total $9+27+9=45$ unknown parameters in the theory. Presence of 
leptoquarks is phenomenologically identical to RPV SUSY. 

For our purposes, the presence of RPV means that 
{\it tree-level} amplitudes become possible in which a virtual 
sparticle propagates from one of the trilinear vertices in 
Eq.~(\ref{rpv1}) to another.  In particular, the FCNC sector probed by charm decays 
involves the $\{ \lambda_{ijk}'\}$. Introducing matrices 
${\cal U}_{\rm L}$, ${\cal D}_{\rm R}$ to rotate left-handed up-quark 
fields and right-handed down-quark fields to the mass basis, we 
obtain for the relevant part of the superpotential 
\begin{eqnarray}
{\cal W}_{\lambda'} = & {\tilde \lambda}_{ijk}' [ 
-\tilde{e}_L^i\bar{d}_R^k u_L^j - \tilde{u}_L^j\bar{d}_R^ke_L^i \\ \nonumber
-& (\tilde{d}_R^k)^*(\bar{e_L^i})^cu_L^j + \dots ] \ \ ,
\label{rpv2}
\end{eqnarray}
where neutrino interactions are not shown and we define 
\begin{equation}
\tilde{\lambda'}_{ijk}\equiv \lambda'_{irs} {\cal U}^L_{rj} 
{\cal D}^{*R}_{sk} \ \ .
\label{rpv3}
\end{equation}
Some bounds on the $\{ \tilde{\lambda'}_{ijk} \}$ are already 
available from data on such diverse sources as 
charged-current universality, the ratio 
$\Gamma_{\pi\to e\nu_e}/\Gamma_{\pi\to \mu \nu_\mu}$, 
the semileptonic decay $D\to K \ell \nu_\ell$ , {\it etc}.

For the decay $D^+\to \pi^+ e^+ e^-$, the effect is proportional to 
${\tilde\lambda}'_{11k}\cdot {\tilde\lambda}'_{12k}$.
Although the effect on the branching ratio is not 
large,  the dilepton spectrum away from 
resonance poles is sensitive to the RPV contributions.  

Another interesting mode is $D^0 \to \mu^+\mu^-$.
Upon using the bound, we obtain 
\begin{equation}
{\cal B}^{\not R_p}_{D^0\to\mu^+\mu^-} < 3.5\times 10^{-6} 
 \ \ .
\end{equation}
A modest improvement in the existing limit on 
${\cal B}_{D^0\to\mu^+\mu^-}$ will be very interesting. 

Lepton flavor violating processes are allowed by the 
RPV lagrangian.  One example is the mode $D^0 \to e^+ \mu^-$, 
for which existing parameter bounds predict 

\begin{eqnarray}
\quad \quad \quad {\cal B}^{\not R_p}_{D^0\to\mu^+e^-} <0.5 \times10^{-6}
\end{eqnarray}

In RPV SUSY, there are also potentially large contributions to several
other 
dilepton modes.  For example, ${\cal B} (D \rightarrow e \nu_\tau)$ can be as
large as $10^{-4}$ and ${\cal B} (D_s \rightarrow e \nu_\tau)$ can be as large
as $5\times 10^{-3}$; these are to be compared to the SM rates for $D
\rightarrow e \nu_e$ of $3\times 10^{-9}$ and $D_s \rightarrow e \nu_e$ of
$1.6\times 10^{-7}$.  There are also enhancements in $D \rightarrow e^+ e^-
\gamma$ and $D^0 
\rightarrow \mu^+ \mu^- \gamma$, 
this time at high $q^2$ and the BR can be enhanced by as much as 20 for the $\mu^+ \mu^- \gamma$ mode.

\begin{table}
\caption{${\cal B}$ for the SM and RPV SUSY (See references 5-8).}
\label{table:1}
\begin{tabular}{|lll|}
\hline
Decay Mode      &       ${\cal B}$ SM  &      ${\cal B}\not{R}_P$ \\
\hline
$D^+ \rightarrow \pi^+e^+e^-$     &    $2 \times 10^{-6}$   &   
$2.3\times 10^{-6}$ \\
$D^+ \rightarrow \pi^+\mu^+\mu^-$ &    $1.9\times 10^{-6}$   & 
$1.5\times 10^{-5}$ \\
$D^0 \rightarrow \rho^0e^+e^-$    &    $1.8\times 10^{-6}$   & 
$5\times 10^{-6}$ \\
$D^0 \rightarrow \pi^0 e^+e^-$    &    $0.8\times 10^{-6}$   &      \\
$D^0 \rightarrow \pi^+ \nu \bar{\nu}$    &     $5\times 10^{-16}$  &        \\
$D^0 \rightarrow \bar{K}^0 \nu \bar{\nu}$    &     $2.4\times 10^{-16}$ &   \\
$D_s \rightarrow \pi^+ \nu \bar{\nu}$     &     $8\times 10^{-15}$  &   \\
$D^0 \rightarrow \gamma \gamma$    &     $(1-3) 10^{-8}$   &  \\
$D^0 \rightarrow \mu^+ \mu^-$     &    $(3-8)10^{-13}$  &   $3\times 10^{-6}$ \\
$D^0 \rightarrow  e^+ e^-$        &    $10^{-23}$     &    $10^{-10}$ \\
$D^0 \rightarrow \mu^\pm e^\mp$    &     0   &  $10^{-6}$    \\
$D^+ \rightarrow \pi^+ \mu^\pm e^\mp$   &   0  &  $3\times 10^{-5}$        \\
$D^0 \rightarrow \rho^0 \mu^+e^-$    &   0  &  $1.4\times 10^{-5}$   \\
$D^0 \rightarrow e^+e^- \gamma$     &  $10^{-9}$ &   $5\times 10^{-9}$   \\
$D^0 \rightarrow \mu^+ \mu^- \gamma$     &  $10^{-9}$   &   $5\times 10^{-8}$ \\
$D^+ \rightarrow  e^+ \nu$        &     $3\times 10^{-9}$   &     $10^{-3}$ \\
$D_s \rightarrow  e^+ \nu$        &     $10^{-7}$     &    $10^{-2}$
\\
\hline
\end{tabular}
\end{table}

\subsection{Extra Scalars and Fermions}

With extra scalar bosons which couple to both quarks and 
leptons, decay modes such as $D^0 \rightarrow \mu^+ \mu^-$ 
and $D^0 \rightarrow \mu^- e^+ + \mu^+ e^-$ can be enhanced.  This is also true of new gauge bosons (e.g. gauged family symmetry).  Extra quarks of charge -1/3 and extra leptons can also perform the same function.  In all these, ${\cal B}$ as high as $10^{-10}$ for $D^0 \rightarrow \mu^+ \mu^-$ and $10^{-13}$ for $D^0 \rightarrow \mu^- e^+ + \mu^+e^-$ can be reached.

\subsection{Large Extra Dimensions}
For several years, the study of large extra dimensions 
(`large' means much greater than the Planck scale) has 
been an area of intense study.  This approach might hold the 
solution of the hierarchy problem while having verifiable 
consequences at the TeV scale or less.  Regarding the subject 
of rare charm decays, one's reaction might be to ask, 
{\it ``How could extra 
dimensions possibly affect the decays of ordinary hadrons?}''
We provide a few examples in the following.

Suppose the spacetime of our world amounts to a $3+1$ brane 
which together with a manifold of additional dimensions 
(the bulk) is part of some higher-dimensional space.  
A field $\Theta$ which can propagate in a large extra 
dimension will exhibit a Kaluza-Klein (KK) tower of 
states $\{ \Theta_n \}$, detection of which would signal existence 
of the extra dimension.  Given our ignorance regarding properties 
of the bulk or of which fields are allowed to propagate 
in it, one naturally considers a variety of different models.

Assume, for example, the existence of an extra dimension of 
scale $1/R \sim 10^{-4}$~eV such that the gravitational field 
(denote it simply as $G$) alone can propagate in the extra 
dimension.  There are then bulk-graviton KK states 
$\{ G_n\}$ which couple to matter.  In principle there will be 
the FCNC $c \to u~G_n$ transitions  and since the $\{ G_n\}$ 
remain undetected, there will be apparent missing energy.  
However this mechanism leads to too small a rate to be observable.

Another possibility which has been studied is that the scale 
of the extra dimension is $1/R \sim 1$~TeV and that SM gauge 
fields propagate in the bulk.  However, precision
electroweak data constrain the mass of the first gauge KK excitation to
be in excess of 4 TeV, and hence their contributions to rare
decays are small.

More elaborate constructions, such as allowing fermion fields to 
propagate in the five-dimensional bulk of the Randall-Sundrum 
localized-gravity model, are currently being actively 
explored.

\subsection{Strong Dynamics}
New strong interactions can be responsible for electro-weak symmetry
breaking and/or the fermion masses.  Some examples are top
condensates, extended technicolor etc.  In general these can lead to FCNC couplings of $Z
\rightarrow cu$ and hence to decay modes such as $D^0 \rightarrow \mu^+ \mu^-$ at a level of $10^{-10}$.

\section{SUMMARY AND CONCLUSIONS}

I would like to think that I have convinced you that there are opportunities in the Q=+2/3 sector.  There are large windows of opportunities for new physics to appear in many modes despite large long distance effects.

In the near term it will be very interesting to improve bounds on modes
such as $D^0 \rightarrow \mu^+ \mu^-, \ 
D \rightarrow \ \rho \ell^+ \ell^-, \ D \rightarrow \pi \ell^+ \ell^-$ and search for $D/D_s \rightarrow e \nu$ and study $D \rightarrow \ell^+ \ell^- \gamma$.

\section{ACKNOWLEDGEMENT}

I would like to thank the organizers, and especially Nick Solomey, for outstanding hospitality and a stimulating meeting.  I thank my collaborators Gustavo Burdman, Gene Golowich, JoAnne Hewett, and also Svetlana Fajfer and Paul Singer for many enjoyable discussions.  This work was supported in part by US.D.O.E under Grant \#DE-FG02-04ER41291.

\end{document}